\documentclass[10pt,journal,twocolumn]{IEEEtran}
\IEEEoverridecommandlockouts
\usepackage{epsfig,cite,bm,algorithm,algorithmic,epstopdf,amsmath,subfigure,multirow,amssymb,amsfonts,amsthm,xcolor,caption,graphicx,textcomp,xcolor, subfigure,array,colortbl}
\def\QEDclosed{\mbox{\rule[0pt]{1.5ex}{1.5ex}}}
\usepackage[T1]{fontenc}
\usepackage{stfloats}
\allowdisplaybreaks[4]

\allowdisplaybreaks[4]
\usepackage{booktabs}

\begin{document}

\title{Joint Beam Scheduling and Resource Allocation for Flexible RSMA-aided Near-Field Communications}

\author{Jiasi Zhou, Cong Zhou, Yijie Mao,~\IEEEmembership{Member,~IEEE}, and Chintha Tellambura,~\IEEEmembership{Fellow,~IEEE}
\thanks{Jiasi Zhou is with the School of Medical Information and Engineering, Xuzhou Medical University, Xuzhou, 221004, China, (email: jiasi\_zhou@xzhmu.edu.cn). (\emph{Corresponding author: Jiasi Zhou}).}
\thanks{Cong Zhou is with the School of Electronic and Information Engineering, Harbin Institute of Technology, Harbin 150001, China, (email: zhoucong@stu.hit.edu.cn).}
\thanks{Yijie Mao is with the School of Information Science and Technology, Shanghai Tech University, Shanghai, 201210, China (email: maoyj@shanghaitech.edu.cn).}
\thanks{ Chintha Tellambura is with the Department of Electrical and Computer Engineering, University of Alberta, Edmonton, AB, T6G 2R3, Canada (email: ct4@ualberta.ca).} 
\thanks{This work was supported by the national key research and development program of China (2020YFC2006600) and the Talented Scientific Research Foundation of Xuzhou Medical University (D2022027).}\vspace{-2.2em}}
\maketitle

\begin{abstract}
Supporting immense throughput and ubiquitous connectivity holds paramount importance for future wireless networks. To this end,  this letter focuses on how the spatial beams configured for legacy near-field (NF) users can be leveraged to serve extra NF or far-field users while ensuring the rate requirements of legacy NF users. In particular, a flexible rate splitting multiple access (RSMA) scheme is proposed to efficiently manage interference, which carefully selects a subset of legacy users to decode the common stream. Beam scheduling, power allocation, common rate allocation, and user selection are jointly optimized to maximize the sum rate of additional users. To solve the formulated discrete non-convex problem, it is split into three subproblems. The accelerated bisection searching, quadratic transform, and simulated annealing approaches are developed to attack them.  Simulation results reveal that the proposed transmit scheme and algorithm achieve significant gains over three competing benchmarks. 
\end{abstract} 


\section{Introduction}
Future wireless networks will evolve towards extremely large-scale antenna arrays and high-frequency spectra to support immense throughput and ubiquitous connectivity\cite{10135096}. Such a trend inevitably expands the Rayleigh distance, which defines the boundary between the near-field (NF) and far-field (FF) regions. The electromagnetic propagation characteristics of NF and FF are markedly distinct, i.e., from planar-wave to spherical-wave\cite{10220205}. The spherical propagation introduces a distance domain, allowing spatial beam energy to be focused on a specific point\cite{10315058}. Nevertheless, the beams designed for different NF users may not be perfectly orthogonal. This provides an opportunity to harness these preconfigured beams (PBs) for supporting additional NF or FF users\cite{10315058}.

References \cite{10315058,10522677}  have exploited PBs to improve throughput and connectivity further. They employ non-orthogonal multiple access (NOMA) to manage interference between legacy and additional users. However, NOMA gains occur only under specific scenarios, such as complex receiver design and significant channel disparity among users\cite{10286271}. An emerging solution is rate splitting multiple access (RSMA), which has better interference management abilities\cite{10038476}. It includes space division multiple access (SDMA) and NOMA as special cases\cite{mao2018rate}. RSMA has already been applied to FF communications \cite{9831440}. This work requires all users to decode the common stream. However, the users with the worst channel determine the common rate,  which hinders the benefits of RSMA\cite{9835151}. RSMA-enabled NF communications remain largely unexplored, except for \cite{10414053}. This work proposes a flexible RSMA scheme and studies the hybrid precoding design, demonstrating that RSMA significantly improves spectral efficiency. To our knowledge, using RSMA  to enhance mixed NF and FF networks has never been considered.  

 To address this gap, this letter examines how spatial PBs for NF users can be utilized to support additional users. A flexible RSMA transmission scheme is proposed, where rate-splitting (RS) users are carefully selected. The approach involves jointly optimizing beam scheduling, power allocation, common rate allocation, and RS-user selection to maximize the sum rate of additional users. This discrete non-convex problem is then decomposed into three subproblems. We develop an accelerated bisection searching (ABS) algorithm to tackle the beam scheduling subproblem. Following this, quadratic transform and simulated annealing approaches are proposed to address the resource allocation (transmit power and common rate) and RS-user selection subproblems. Simulations show that the proposed transmit scheme and algorithm yield significant gains over several baselines.
\section{System model and problem formulation}\label{Section II}
\begin{figure}[tbp]
\centering
\includegraphics[scale=0.65]{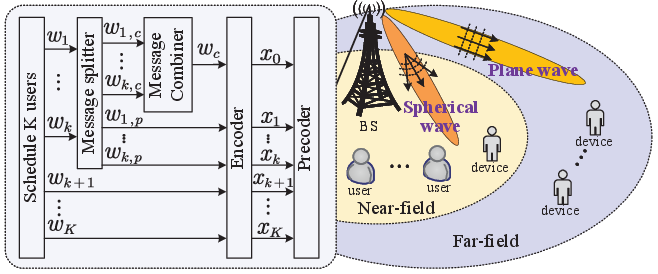}
\caption{Flexible RSMA-enabled NF-communication system.}
\vspace{-0.4cm}
\label{fig:system}
\end{figure} The downlink RSMA-aided multi-user network, Fig.~\ref{fig:system},  comprises an $N$-antenna base station (BS) and $K$ single-antenna NF users, where $N=2\tilde N+1$.  The BS employs a uniform linear array (ULA) with an antenna spacing of $d$. This letter aims to support $M$ additional single-antenna users by leveraging PBs\footnote{ The key difference between our work and \cite{10659020} is that we do not focus on optimizing beamforming and expanding converge.}. For simplicity, the legacy NF users\footnote{Legacy users are users served by dedicated spatial beams to meet their communication requirements.}  are referred to as \emph{users} and the extra NF or FF users as \emph{devices}. Each beam can only accept one device to ensure each user has sufficient performance. These devices may be located in NF or FF regions.  The boundary between these two regions is determined by the Rayleigh distance $Z = \frac{2D^2}{\lambda}$\cite{10220205}, where $D=(N-1)d$ and $\lambda$ are the antenna array aperture and carrier wavelength, respectively.
\vspace{-0.5cm}
\subsection{NF and FF channel models}
Define the coordinate of the ULA's midpoint as $(0,0)$ and set the $y$-axis along the ULA, so the $n$-th antenna is $\mathbf{s}_n=\left(0, \tilde nd\right)$, where $\tilde n = n-\frac{N+1}{2}$. Let $d_k$ and $\theta_k$ denote the distance and angle of user $k$, so its coordinate is $\mathbf{d}_k=\left(d_k\cos\theta_k,d_k\sin\theta_k\right)$. Then, the distance between the $n$-th antenna and this user is calculated as
\begin{equation}
d_{k,n}=||\mathbf{d}_k-\mathbf{s}_n|| = \sqrt{d^2_k+(\tilde nd)^2-2\tilde ndd_k\sin\theta_k}.
\label{distance}
\end{equation}
According to spherical propagation model\cite{10458958}, the NF channel $\mathbf{h}_k\in\mathbb{C}^{N\times 1}$ between the BS and user $k$ is 
\begin{equation}
\mathbf{h}_k= \beta_k\big[e^{-j\frac{2\pi}{\lambda}d_{k,1}},\dots,e^{-j\frac{2\pi}{\lambda}d_{k,N}}\big]^T=\beta_k\mathbf{a}\left(d_k,\theta_k\right).
\label{Channel}
\end{equation}
where $\mathbf{a}\left(d_k,\theta_k\right)$ is NF array response vector. $\beta_k=\frac{c}{4\pi fd_k}$ is the free space path loss, where $f$ and $c$ denote the carrier frequency and speed of light, respectively. Similar to \cite{10315058,10522677}, this letter assumes that perfect channel state information (CSI) is available. The NF array response vector $\mathbf{a}\left(d_k,\theta_k\right)$ downgrades to the FF one when $D \ll d_k$ and by using the first-order Taylor approximation to $d_{k,n}$\cite{10135096}. This case yields  $d_{k,n}\approx d_k-\tilde nd\sin\theta_k$. Plugging the approximated $d_{k,n}$ into equation (2), the FF channel $\tilde{\mathbf{h}}_{far}\in\mathbb{C}^{N\times 1}$ can be modelled as  
\begin{equation}
\tilde{\mathbf{h}}_{far}= \tilde\beta\big[e^{-j\frac{2\pi}{\lambda}\tilde Nd\sin\tilde\theta},\dots,e^{j\frac{2\pi}{\lambda}\tilde Nd\sin\tilde\theta}\big]^T=\tilde\beta\tilde{\mathbf{a}}\left(\tilde d,\tilde\theta\right), 
\label{Far-Channel}
\end{equation}
where $\tilde\beta=\frac{c}{4\pi f\tilde d} e^{-j\frac{2\pi}{\lambda}\tilde d}$, and $(\tilde d,\tilde\theta)$ is the corresponding coordinate.
Consequently, the channel $\mathbf{g}_m\in\mathbb{C}^{N\times 1}$ between the BS and the $m$-th device is
\begin{equation}\label{fra}
\mathbf{g}_m=
\begin{cases}
\beta_m\mathbf{a}\left(d_m,\theta_m\right),&\mbox{if $d_m\leq Z$ };\\
\tilde\beta_m\tilde{\mathbf{a}}\left(d_m,\theta_m\right), &\mbox{if $d_m>Z$ }.
\end{cases}
\end{equation}

\vspace{-0.5cm}
\subsection{Signal model and problem formulation}
Unlike conventional RSMA, our scheme allows a subset of users to decode the common stream via successive interference cancellation (SIC). For convenience, $K$ users are divided into two separate groups, indexed by $\mathcal{K}_1$ and $\mathcal{K}_2$, where users in $\mathcal{K}_1$ and $\mathcal{K}_2$ decode and do not decode the common stream, respectively. We introduce a binary variable $s_k$, where $s_k=1$ indicates $k\in\mathcal{K}_1$, otherwise $s_k=0$. The message for user $k$ in $\mathcal{K}_1$ is split into common and private parts. All common parts are encoded to one common stream $x_0$ while the private part for user $k$ is encoded to $x_k$. The message for user $k'$ in $\mathcal{K}_2$ is directly encoded into a private stream $x_{k'}$. These streams are mutually independent and precoded by $\mathbf{p}_0\in\mathbb C^{N\times 1}$ and $\mathbf{p}_k\in\mathbb C^{N\times 1}$. Herein, the zero-forcing precoder is adopted for private streams, i.e., $[\mathbf{p}_1,\dots,\mathbf{p}_K] = \mathbf{H}\left(\mathbf{H}^H\mathbf{H}\right)^{-1}\mathbf{F}$, where $\mathbf{H}=[\mathbf{h}_1,\dots,\mathbf{h}_K]$ and $\mathbf{F}\in\mathbb{C}^{N\times N}$ is a diagonal matrix to ensure power normalization with
$[\mathbf{F}]_{n,n} =\big[\left(\mathbf{H}^H\mathbf{H}\right)^{-1}\big]^{-\frac{1}{2}}_{n,n}$.
Additionally, since the common rate depends on the user with the worst channel quality, we set $\mathbf{p}_0=\mathbf{h}_{k'}||\mathbf{h}_{k'}||^{-1}$, where $k'=\arg\min_k\left\{||\mathbf{h}_1||,\dots,||\mathbf{h}_K||\right\}$. This letter focuses on how to use these PBs to serve additional devices. 

Let binary variable $b_{k,m}=1$ if the $m$-th device occupies the $k$-th beam, otherwise $b_{k,m}=0$. Therefore, the transmitted signal is 
$\mathbf{x}=\sum_{k=0}^{K}\mathbf{p}_{k}\left(\sqrt{P_k}x_k +\sum_{m=1}^{M}b_{k,m}\sqrt{P_{k,m}}\tilde x_m \right)$, where
$P_k (P_0)$ and $P_{0,m} (P_{k,m})$ are the power allocated to the user and device on the $k$-th beam (common stream's beam).
The received signal at user $k$ is $y_k=\mathbf{h}^H_k\mathbf{x} + n_k$ for $ \forall k\in\mathcal{K}=\{1,\dots,K\}$, where $n_k\sim \mathcal{CN}\left(0,\sigma^2\right)$ is the additive white Gaussian noise (AWGN). To retrieve the desired message, user $k$ in $\mathcal{K}_1$ decodes the common stream by treating other streams as noise. Based on the zero-forcing principle,  the signal-to-interference-plus-noise ratio (SINR) is
\begin{equation}
\gamma_{k,c}=\frac{h_{k,0}P_0}{h_{k,k}P_k + I_k +  \sigma^2}, \quad \forall k\in\mathcal{K}_1,\label{common_rate}
\end{equation}
where $I_k=\sum_{m=1}^{M} \left(h_{k,0}b_{0,m}P_{0,m}+h_{k,k}b_{k,m}P_{k,m}\right)$ and $h_{k,i}={\left|\mathbf{h}^H_{k}\mathbf{p}_i\right|}^2$.
To ensure all users in $\mathcal{K}_1$ can recover $x_0$, the common rate shall not exceed $R_c= \min_{\forall k\in\mathcal{K}_1}\log\left(1+\gamma_{k,c}\right)$. Moreover, all users in $\mathcal{K}_1$ share the common rate, so  $\sum_{k=1}^{K}s_kR_{k,c} \leq R_c$, where $R_{k,c}$ is the common rate allocated to the $k$-th user. After removing $x_0$, users in $\mathcal{K}_1$ decode their private stream. In contrast, users in $\mathcal{K}_2$ directly decode their desired streams by treating other streams as noise. To save space,  we merge the SINRs of users in $\mathcal{K}_1$ and $\mathcal{K}_2$ decoding private streams into one expression, which is given as 
\begin{equation}
\gamma_{k,p}=\frac{h_{k,k}P_k}{(1-s_{k})h_{k,0}P_0 + I_k + \sigma^2_k},\quad k\in\mathcal{K}.\label{private_common}
\end{equation}
Therefore, the $k$-th user's rate is $R_k=s_k R_{k,c}+\log\left(1+\gamma_{k,p}\right)$.
Similarly, the received signal of the $m$-th device is $y_m=\mathbf{g}^H_m\mathbf{x} + n_m$, where $n_m\sim \mathcal{CN}\left(0,\sigma^2\right)$ is AWGN. Assuming that $k$-th beam support the $m$-th device, the corresponding SINR is 
\begin{equation}
\tilde \gamma_{m}=\frac{g_{k,m}P_{k,m}}{\sum_{i=0}^{K}g_{i,m}\left(P_i +\sum_{j=1,j\neq m}^{M}b_{i,j}P_{i,j}\right)  + \sigma^2}.\label{additional-rate}
\end{equation}
where $g_{k,m}={\left|\mathbf{g}^H_{m}\mathbf{p}_k\right|}^2$. As a result, the achievable rate of the $m$-th device is $R_{m}=\log\left(1+\tilde \gamma_{m}\right)$.

This letter aims to maximize the sum rate of devices while ensuring the rate requirement of users by jointly optimizing beam scheduling, power allocation, common rate allocation, and RS-user selection. This problem is formulated as
\begin{subequations}\label{linear_p}
	\begin{align}
&\max_{P_{k},P_{k,m},R_{k,c},s_k,b_{k,m}} \sum_{m=1}^{M}R_{m},\label{ob_a}\\
	\text{s.t.}~
	&\sum_{k=0}^{K} \left(P_k +\sum_{m=1}^{M}b_{k,m}P_{k,m} \right)\leq P_{max},\label{ob_b}\\&
 R_k\geq R_{th}, \quad \forall k,\label{ob_c}\\
 &\sum_{k=1}^{K}s_kR_{k,c} \leq R_c,\label{ob_d}\\
 &R_{k,c} \geq 0, \quad \forall k\in\mathcal{K}_1,\label{ob_e}\\
  &\sum_{m=1}^{M}b_{k,m}=\sum_{k=1}^{K}b_{k,m}= 1, b_{k,m}\in\{0,1\},\forall k,m\label{ob_f}\\
        &s_{k}\in\{0,1\},~\forall k,\label{ob_g}
	\end{align}
\end{subequations}
where $P_{max}$ and $R_{th}$ are maximum transmit power and minimum rate requirement thresholds, respectively. 

Problem (\ref{linear_p}) is highly complex, imposing three technical challenges. First, the optimal beam scheduling and RS-user selection require exhaustive searching. This is prohibitive when many beams or users are scheduled. Second, the logarithmic function in decoding streams is non-convex. Due to the unknown duality gap, such problems are difficult to solve in the primal and dual domains. Third, the minimum operator in $R_c$ introduces non-smoothness, complicating resource allocation. Consequently, the optimal solution appears incomputable.

\section{Algorithm design and properties analysis}
This section decomposes problem (\ref{linear_p}) into three subproblems: 1) a beam scheduling; 2) a power and common rate allocation; and 3) a binary RS-user selection. An ABS algorithm is developed to address the first one. The quadratic transform and simulated annealing approaches are utilized to solve the last two subproblems.
\vspace{-0.5cm}
\subsection{Beam scheduling subproblem} 
This problem is solved with the following proposition. 

\noindent\emph{\textbf{Proposition 1:}} The transmit rates of users are independent of beam scheduling while device $m$ prefers beam $k'$ than beam $k$ if $g_{k',m}>g_{k,m}$.

\emph{Proof}: Equations (\ref{common_rate}) and (\ref{private_common}) indicate that the corresponding SINRs depend on power allocation and RS-user selection. However, it is independent of the beam scheduling when the power allocation is identical, so SINRs remain static no matter how the beam is scheduled. Similarly, equation (\ref{additional-rate}) shows that the transmission rate of device $m$ increases with $g_{k,m}$. Therefore, to boost transmit rate, device $m$ prefers beam $k'$ with $k'=\arg\max_{\forall k} g_{k,m}$. This proves Proposition 1.\hfill \QEDclosed

Proposition 1 reveals the beam scheduling principle, but multiple devices may expect to occupy one beam. To attack this challenge, we recast the beam scheduling problem as 
\begin{subequations}\label{linear_p2}
	\begin{align}
&\max_{b_{k,m}}\min g_{k,m},\label{ob_a2}\\
	\text{s.t.}~
	&\mbox{ (\ref{ob_f})}
	\end{align}
\end{subequations}
The optimized objective can avoid significant performance degradation in some beams. To help solve the problem (\ref{linear_p2}), construct matrix $\mathbf{G}=[\tilde{\mathbf{g}}_1,\dots,\tilde{\mathbf{g}}_M]$, where $\tilde{\mathbf{g}}_m=[g_{1,m},\dots,g_{K,m}]^T$. Next,  an ABS algorithm is proposed  with searching range $[\alpha_{\min},\alpha_{max}]$, where $\alpha_{\min}=\min_{\forall m,k}g_{k,m}$ and $\alpha_{\max}=\min_{\forall m}\max_{\forall k}g_{k,m}$. Given a required threshold $\epsilon_{th}$, this algorithm first determines whether $g_{k,m}\geq \epsilon_{th}$ can be satisfied for each beam. Then, according to the scheduling result, $\epsilon_{th}$ is updated based on the bisection searching approach. Two definitions are introduced to help describe the procedure.

\emph{\textbf{Definition 1:}} If $g_{k,m}\geq\epsilon_{th}$, $k$ is called a feasible beam of device $m$ while $m$ is an alternate device of beam $k$. The number of feasible beams is the degree, denoted by $\mathrm{deg}(m)$.

\emph{\textbf{Definition 2:}} The scheduling priority of the $m$-th device is higher than that of $m'$ if

1) $\mathrm{deg}\left(m\right)<\mathrm{deg}\left(m'\right)$, or

2) $\mathrm{deg}\left(m\right)=\mathrm{deg}\left(m'\right)$ and $m<m'$.\\
Condition 2) helps to avoid two devices with the same degree. Similarly, define $\mathrm{deg}(k)$ and its priority. The reasons for definition 2 are as follows. If device $m$ is with $\mathrm{deg}\left(m\right)=0$, it cannot occupy any beam since there is no $k$ that can meet $g_{k,m}>\epsilon_{th}$. Thus,  higher priority should be endowed to the device with a lower degree. Meanwhile, it selects the beam with the highest priority.

Alg.~\ref{Alg.1} summarizes the scheduling procedures. Lines 2$\sim$8  judge whether $\epsilon_{th}$ can be reached. Specifically, line 7 updates $g_{x,y}$ to ensure that already scheduled beams no longer interfere with subsequent matching.  To accelerate the convergence rate, the searching interval length must be narrowed in every loop (lines 12 and 15). For example, in line 12, if $\mathrm{deg}\left(m\right)=1$, the minimum has been maximized and cannot be improved. Thus, the algorithm prematurely terminates (Lines 12$\sim$14).

\begin{algorithm}[t]
	\caption{Accelerated Bisection Searching Algorithm }
	\begin{algorithmic}[1]\label{Alg.1}
		\STATE Calculate matrix $\mathbf{G}$, initiate  $\alpha_{\min}$, $\alpha_{\max}$, iteration index $i=1$, the maximum tolerance $\epsilon_{th}$, and $\epsilon^{(1)}-\epsilon^{(0)}>\epsilon_{th}$.
		\WHILE {$|\epsilon^{(i)}-\epsilon^{(i-1)}|>\epsilon_{th}$}
		\STATE Set $\epsilon^{(i)}=0.5(\alpha_{\min}+\alpha_{\max})$, $\mathcal{S}=\tilde{\mathcal{S}}=\emptyset$, and $\mathbf{g} = \mathbf{0}_{M}$.
        \STATE Set $g_{k,m}=0$ for $\forall g_{k,m}\leq \epsilon^{(i)}$.
		\WHILE{ $\mathrm{deg}\left(m\right) \neq 0$ for $\forall m\in\mathcal{M}\backslash\tilde{\mathcal{S}}$} 
		\STATE Search $m$ with the highest priority and then search $k$ with the highest priority from its neighbor set.
		\STATE Update $\mathcal{S}= \mathcal{S}\cup (m,k)$, $\tilde{\mathcal{S}}=\mathcal{S}\cup m$, $g_{x,y}=0$ for $x=k$ or $y=m$, $\mathbf{g}\left(j\right)=g_{k,m}$, and $j=j+1$.
		\ENDWHILE
		\STATE Update $i=i+1$;
		\IF {$j=M$}
        \STATE Set $g_{k,m}=0$ for $\forall g_{k,m}\leq \min\left(\mathbf{g}\right)$.
		\IF{ $\mathrm{deg}\left(m\right)=1$, where $m=\arg\min_{g_{k,m}} \mathbf{g}$ }
		\STATE break;
		\ENDIF	
		\STATE Update $\alpha_{\min}=\min\left(\mathbf{g}\right) \geq \epsilon^{(i-1)}$.
		\ELSE
		\STATE Update $\alpha_{\max}=\epsilon^{(i-1)}$.
		\ENDIF
		\ENDWHILE
		\STATE Output beam scheduling result  $\mathcal{S}$.
	\end{algorithmic}
\end{algorithm}

\vspace{-0.5cm}
\subsection{Power and common rate allocation subproblem}
With the known beam scheduling and RS-user selection,  fractional SINR in problem (\ref{linear_p}) hampers optimization. To address this challenge, surrogate optimization is utilized, which involves constructing easily optimizable surrogates for complex constraints. To minimize performance loss, these surrogates must be strictly equivalent or closely approximated to the primary function. To this end, accurate surrogates are constructed using a quadratic transform approach \cite{shen2018fractional}. Theorem 2 in \cite{shen2018fractional} motivates Claim 1.

{\bf\emph{Claim 1}}: For function $f\left(y,p\right) = 2y\sqrt{s(p)}-y^2 I(p)$ for any $s(p)\geq 0$ and $I(p)> 0$, one has $
s(p)I^{-1}(p) =\max_{y>0}f\left(y,p\right)$.
The optimal solution to  $\max_{y>0}f\left(y,p\right)$ is $y^*=\sqrt{s(p)}I^{-1}(p)$.

\emph{Proof}: Please see \cite{shen2018fractional} for the detailed proof.\hfill \QEDclosed

Now, Claim 1 is engaged  to reformulate $\gamma_{k,c}$, $\gamma_{k,p}$, and $\tilde{\gamma}_m$. The constructed surrogates can be written as \begin{subequations}\label{surrogate}
	\begin{align}
&f\left(\hat{\mathbf{p}},y_{k}\right)=2y_{k}\sqrt{h_{k,0}P_0}-y^2_{k}\left(h_{k,k}P_k + I_k +  \sigma^2\right),\\
 &f\left(\hat{\mathbf{p}},\hat y_{k}\right)=2\hat y_{k}\sqrt{h_{k,k}P_k}-\hat y^2_{k}\left((1-s_k)h_{k,0}P_0 + I_k+ \sigma^2\right),\\
 &f\left(\hat{\mathbf{p}},\tilde y_{m}\right)=2\tilde y_{m}\sqrt{g_{k,m}P_{k,m}}-\tilde y^2_{m}\left(\tilde I_m + \sigma^2\right),
	\end{align}
\end{subequations}
where $\tilde I_m = \sum_{i=0}^{K}g_{i,m}\left(P_i +\sum_{j=1,j\neq m}^{M}b_{i,j}P_{i,j}\right)$ and $\hat{\mathbf{p}}=\{P_0,\dots,P_K,P_{0,1},\dots,P_{K,M}\}$.
After removing the minimum operator in (\ref{ob_d}), this subproblem can be recast as
\begin{subequations}\label{linear_p3}
	\begin{align}
&\max_{\mathcal{Q}_1,\mathcal{Q}_2} \sum_{m=1}^{M}R_{m},\label{ob_a3}\\
	\text{s.t.}~
 &s_kR_{k,c}+\log\left(1+f\left(\hat{\mathbf{p}},\hat y_{k}\right)\right)\geq R_{th},\forall k\in\mathcal{K},\label{ob_b3}\\
 &\sum_{k=1}^{K}s_kR_{k,c} \leq \log\left(1+f\left(\hat{\mathbf{p}},y_{k}\right)\right),\forall k\in\mathcal{K}_1,\label{ob_c3}\\
  & \log\left(1+f\left(\hat{\mathbf{p}},\tilde y_{m}\right)\right)\geq R_{m},\forall m\in\mathcal{M},\label{ob_d3}\\
 &\mbox{(\ref{ob_b}), (\ref{ob_e})}
	\end{align}
\end{subequations}
where $\mathcal {Q}_1=\{P_{k},P_{k,m}, R_{k,c}\}$ and $\mathcal {Q}_2=\{y_{k},\hat y_{k},\tilde y_m\}$. It is observed that problem (\ref{linear_p3}) becomes convex upon fixing $\mathcal{Q}_2$. Additionally, once  $\mathcal{Q}_1$ are specified, the optimal $\mathcal{Q}_2$ can be derived with closed-form expressions via Claim 1. Given this observation, problem (\ref{linear_p3}) is recast as a two-tier alternating optimization problem,  where $\mathcal {Q}_1$ and $\mathcal {Q}_2$ are optimized alternately by the convex framework and closed-form expressions, respectively. The solution process is summarized in Alg.~2.
\vspace{-0.5cm}
\subsection{Binary RS-user selection subproblem}
The solution space of RS-user selection is in the order of $\mathcal{O}\left(2^K\right)$. The optimal solution requires an exhaustive search, incurring unaffordable computational complexity. To reduce complexity, this letter proposes a simulated annealing-based algorithm to find an efficient solution, which can escape from local optima via exploration\cite{10414053}. Thus, the overall algorithm for the problem (\ref{linear_p}) is summarized as Alg.~\ref{Alg.3}. Specifically, in each iteration, one of the binary variables from $\mathbf{s}=\{s_1,\dots,s_k,\dots,s_K\}$ is changed to $\tilde{\mathbf{s}}=\{s_1,\dots,1-s_k,\dots,s_K\}$. Then,  whether this change can increase the sum rate is determined. If $R(\tilde{\mathbf{s}})>R(\mathbf{s})$ holds,  $s_k$ is changed to $1-s_k$, where $R(\mathbf{s})$ denotes sum rate under the RS-user selection result $\mathbf{s}$ via Alg.~\ref{Alg.2}.  Otherwise, the change $s_k$ to $1-s_k$ is made with probability 
\begin{equation}
\mathbb{P}=\exp\left(\frac{R(\tilde{\mathbf{s}})-R(\mathbf{s})}{\delta}\right)\in(0,1]. 
\label{Probability}
\end{equation} 
The parameter $\delta$ controls the percentage of changing $s_k$ to $1-s_k$. Its value is initially set relatively large and then decreases gradually as the number of iterations increases. Several crucial properties of Alg.~\ref{Alg.3} are discussed next.

\emph{Convergence and optimality:} Given $\epsilon_{th}$, if there is a beam scheduling scheme that can ensure $g_{k,m}\geq \epsilon_{th}$ for $\forall m$, the inner of Alg.~\ref{Alg.1} ends after $M$ searchings, otherwise, it terminates prematurely. Besides, without the acceleration strategy, the feasible interval is reduced by half after each iteration, leading to  ${\left(\frac{1}{2}\right)}^i\leq \epsilon_{th}$. Therefore, the inner and outer loops of Alg.~\ref{Alg.1} will terminate and output a deterministic beam scheduling result after $M$ and $\lceil-\log_2\left(\epsilon_{th}\right)\rceil$ iterations in the worst-case, respectively, where $\lceil\rceil$ denotes an integer up.  Alg. ~\ref{Alg.2} yields the optimal solutions in lines 3 and 4. This indicates that it can search the last feasible point at least, so it produces a non-decreasing objective value after each iteration. Moreover, since the sum rate cannot increase infinitely, it is thus deduced that Alg.~\ref{Alg.2} converges to a local optimum, at least within finite iterations. Alg.~\ref{Alg.3} mimics the annealing process by iteratively exploring solutions, facilitating convergence to a local optimum at least\cite{10414053}.

\emph{Complexity:} In Alg.~\ref{Alg.1}, the computational complexity for calculating $g_{k,m}$ is $\mathcal{O}\left(N^2\right)$ while the number of calculations is $MK$. Additionally, the complexity of the inner loop is $\mathcal{O}\left(M^3\right)$ while the outer bisection searching terminates after at most $\lceil-\log_2\left(\epsilon_{th}\right)\rceil$ iterations. Thus, the complexity of Alg.~\ref{Alg.1} is $\mathcal{O}\left(N^2MK+\lceil-\log_2\left(\epsilon_{th}\right)\rceil M^3\right)$. As for Alg.~\ref{Alg.2}, the computational complexity of the conventional interior point method is $\mathcal O(\epsilon_1K^{3.5})$, where $\epsilon_1$ is the number of iterations. Therefore, the complexity of our overall algorithm is $\mathcal O(N^2MK+\lceil-\log_2\left(\epsilon_{th}\right)\rceil M^3 + \epsilon_1\epsilon_2 K^{3.5})$, where $\epsilon_2$ is the number of iterations required by Alg.~\ref{Alg.3}.

\begin{algorithm}[t]
	\caption{Quadratic transform-based iteration algorithm}
	\begin{algorithmic}[1]\label{Alg.2}
        \STATE Initialize $\mathcal{Q}_1^{(1)}$, $s_k$, and iteration index $i=1$.
        \WHILE{ not convergent}
		\STATE  Calculate $\mathcal{Q}_2^{(i)}$ under fixed $\mathcal{Q}_1^{(i)}$ based on Claim 1.
		\STATE  Solve problem (\ref{linear_p3}) under fixed $\mathcal{Q}_2^{(i)}$  and  output $\mathcal{Q}^{(i+1)}_1$.
		\STATE Update iteration index $i=i+1$.
		\ENDWHILE	
		\STATE Output power allocation and sum rate.
	\end{algorithmic}
\end{algorithm}

\begin{algorithm}[t]
	\caption{Overall algorithm for solving problem (\ref{linear_p})}
	\begin{algorithmic}[1]\label{Alg.3}
		\STATE Obtain beam scheduling result by Alg.~\ref{Alg.1}.
        \STATE Initialize $\mathbf{s}^{(0)}=\{1,\dots,1\}$, optimal sum rate $R^*=0$ and iteration index $t=0$. Obtain the initialized sum rate $R^o=R\left({\mathbf{s}}^{(0)}\right)$ via Alg.~\ref{Alg.2}.
        \WHILE{ not convergent}
		\STATE  Obtain $\tilde{\mathbf{s}}^{(t)}$ from  $\mathbf{s}^{(t)}$ by changing $s_k$ to $1-s_k$, where $k={\rm{mod}}\left(t,K\right)+1$ and calculate $R\left(\tilde{\mathbf{s}}^{(t)}\right)$ via Alg.~\ref{Alg.2}.
		\IF{ $R\left(\tilde{\mathbf{s}}^{(t)}\right)>R^o$ }
		\STATE Update ${\mathbf{s}}^{(t+1)}=\tilde{\mathbf{s}}^{(t)}$ and $R^o = R\left(\tilde{\mathbf{s}}^{(t)}\right)$.
        \ELSE
        \STATE Update ${\mathbf{s}}^{(t+1)}=\tilde{\mathbf{s}}^{(t)}$ and $R^o=R\left(\tilde{\mathbf{s}}^{(t)}\right)$ with  probability $\mathbb{P}$. Keep ${\mathbf{s}}^{(t+1)} ={\mathbf{s}}^{(t)}$ with probability $1-\mathbb{P}$.
		\ENDIF
        \IF{$R\left(\tilde{\mathbf{s}}^{(t)}\right)>R^*$ }
        \STATE Set $\mathbf{s}^*=\tilde{\mathbf{s}}^{(t)}$ and $R^*=R\left(\tilde{\mathbf{s}}^{(t)}\right)$
        \ENDIF
		\STATE Update $t=t+1$ and $\delta = \rho\delta$.
		\ENDWHILE
		\STATE Output the optimal RS-user selection and sum rate. 
	\end{algorithmic}
    
\end{algorithm}

\section {Simulation results}\vspace{-0.1cm}
The proposed transmit scheme and algorithm are evaluated next. The circular coverage area has a radius of $120$~m. Users and devices are randomly distributed throughout the NF region and the whole area. Unless specified otherwise, $N=127$, $K=8$, $M=8$, $R_{th}=0.8$~bps/Hz, $P_{max}=30$~dBm, $f_c=30$~GHz, $\lambda=0.01$~m, $\sigma^2=-80$~dBm, $d=\frac{\lambda}{2}$, $Z=\frac{2D^2}{\lambda}=79.4$~m, $\delta=20$, and $\rho=0.9$. These parameters are primarily sourced from \cite{10315058,10522677,10414053,10658996}. 
For comparative performance evaluation,  our approach (legend \emph{FRS-ABS}) is benchmarked against three baselines: (1) All users decode the common stream (legend \emph{RS-ABS}). (2) The BS utilizes SDMA to encode and transmit messages for users (legend \emph{SDMA-ABS}). (3) The additional users randomly occupy the PBs  (legend \emph{FRS-Random}). 

\begin{figure*}[tbp]
	\centering
	\subfigure{
		\begin{minipage}[b]{0.235\linewidth}
		\centering
		\setcounter{figure}{1}	\includegraphics[scale=0.4]{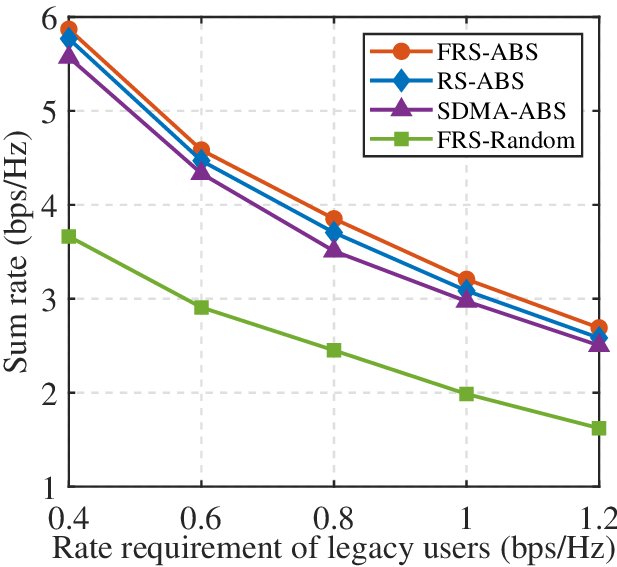}
		\caption{ Sum rate versus the minimum rate requirement.}
		\label{fig:power_LU}
		\end{minipage}
	}
	\subfigure{
		\begin{minipage}[b]{0.23\linewidth}
		\centering
   	\includegraphics[scale=0.4]{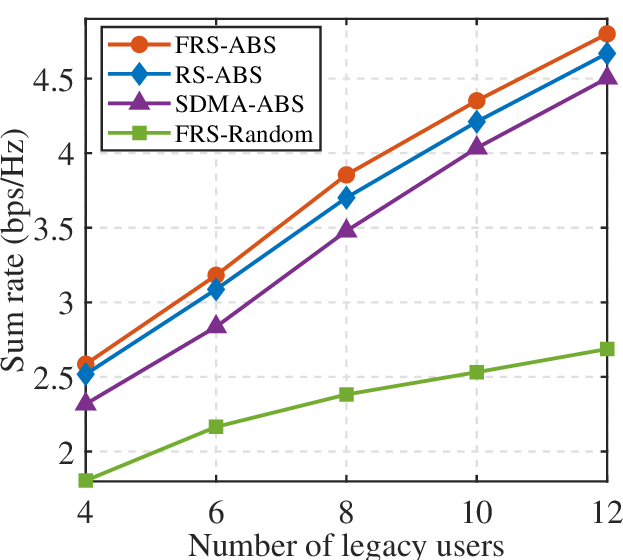}
		\caption{Sum rate versus the number of legacy users.}
		\label{fig:number_LU}
		\end{minipage}
	}
 	\subfigure{
		\begin{minipage}[b]{0.23\linewidth}
		\centering
   	\includegraphics[scale=0.4]{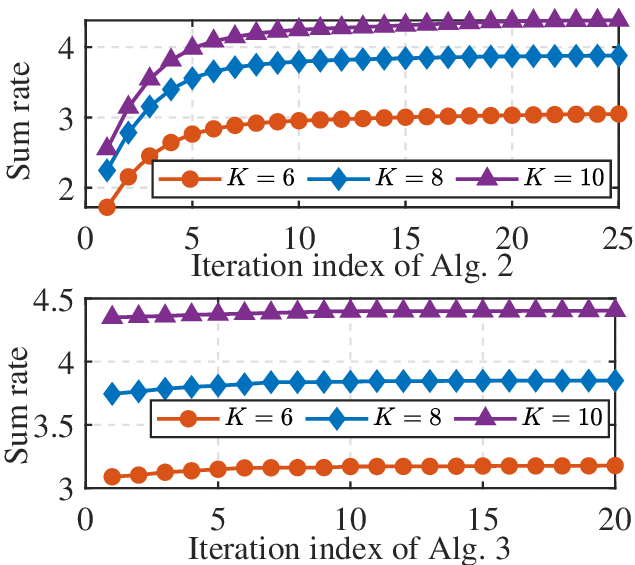}
		\caption{Sum rate versus iteration index.} 
		\label{fig:convergence}
		\end{minipage}
	}
	\centering
 \vspace{-0.6cm}
\end{figure*}

Fig.~\ref{fig:power_LU} plots the sum rate versus minimum rate requirement thresholds, highlighting three key advantages of our approach over the benchmarks. First, RSMA outperforms SDMA, gradually increasing the performance gap and showcasing RSMA's superior interference management. Second, our transmission scheme surpasses conventional RSMA by allowing certain users to bypass decoding common streams and optimizing the RS-user set through the simulated annealing approach. Third, our ABS algorithm outperforms random scheduling, achieving a $55\%$ gain through more effective beam scheduling.

Fig.~\ref{fig:number_LU} illustrates the sum rate versus the number of users. An interesting observation is that for all schemes and algorithms, as the number of legacy users increases, so does the sum rate. This is primarily because the network can support additional users when more legacy users are scheduled. However, our transmission scheme and algorithm are always superior to the three benchmarks, especially when $K\leq 8$. Moreover, the performance gap between our ASB algorithm and beam random scheduling approach becomes more noticeable as the number of users increases. And that again underlines the importance of reasonable scheduling of beams.

Fig.~\ref{fig:convergence} shows the convergence performance, with {\bf Algorithm 3} converging after a sequence of non-decreasing values.  It validates the convergence analysis in Section III.    
\vspace{-0.5cm}
\section{Conclusion}\vspace{-0.1cm}This letter proposes a novel transmit scheme to enhance the throughput and connectivity of RSMA NF networks. The sum rate maximization for extra users presents a discrete non-convex problem, which is challenging to solve. To address this, the problem is divided into three subproblems, each tackled with specialized algorithms. The overall algorithm offers significant gains.  For example,  it achieves considerable gains over traditional RSMA, SDMA, and random beam scheduling. Future research directions encompass exploiting other precoding techniques like maximum ratio transmission to utilize these preconfigured spatial beams and expanding the proposed algorithm to accommodate imperfect CSI.

	\ifCLASSOPTIONcaptionsoff
	\newpage
	\fi

	\bibliographystyle{IEEEtran}
	\bibliography{references}
	
\end{document}